\documentclass[english]{revtex4}
\usepackage{float}
\usepackage{graphicx}
\usepackage{amssymb}

\usepackage[hyperref]{hyperref}

\begin{document}

\title{DVCS via color dipoles: Nonperturbative effects}

\author{B.Z.~Kopeliovich}

\email{Boris.Kopeliovich@usm.cl}

\affiliation{Departamento de F\'{i}sica y Centro de Estudios Subatomicos, Universidad T\'{e}cnica Federico Santa Mar\'{i}a, Casilla 110-V, Avda. Espa\~na 1680, Valpara\'{i}so, Chile}

\author{Ivan Schmidt}

\email{Ivan.Schmidt@usm.cl}

\affiliation{Departamento de F\'{i}sica y Centro de Estudios Subatomicos, Universidad T\'{e}cnica Federico Santa Mar\'{i}a, Casilla 110-V, Avda. Espa\~na 1680, Valpara\'{i}so, Chile}

\author{M. Siddikov}

\email{Marat.Siddikov@usm.cl}

\affiliation{Departamento de F\'{i}sica y Centro de Estudios Subatomicos, Universidad T\'{e}cnica Federico Santa Mar\'{i}a, Casilla 110-V, Avda. Espa\~na 1680, Valpara\'{i}so, Chile}

\date{December 19, 2008}

\begin{abstract}
We study the DVCS amplitude within the color dipole approach. The
light-cone wave function of a real photon is evaluated in the instanton
vacuum model. Our parameter free calculations are able to describe
H1 data, both the absolute values and the $t$-dependences, at medium-high
values of $Q^{2}$. The $Q^{2}$ dependence is found to be sensitive
to the choice of the phenomenological cross section fitted to DIS
data. 
\end{abstract}
\maketitle

\section{Introduction}

During the last decade hard exclusive reactions, such as deeply virtual
compton scattering (DVCS), $\gamma^{*}+p\to\gamma+p,$ have been a
subject of intensive theoretical and experimental investigation~\cite{Mueller:1998fv,Ji:1996nm,Ji:1998pc,Radyushkin:1996nd,Radyushkin:1997ki,Radyushkin:2000uy,Ji:1998xh,Collins:1998be,Collins:1996fb,Brodsky:1994kf,Goeke:2001tz,Diehl:2000xz,Belitsky:2001ns,Diehl:2003ny,Belitsky:2005qn}.
Particular interest deserves the Bjorken kinematics,\begin{eqnarray}
-q^{2} & \equiv & W^{2}Q^{2}\gg\Lambda_{QCD}^{2},\qquad W^{2}\equiv\left(p+q\right)^{2}\gg\Lambda_{QCD}^{2},\\
x_{B} & \equiv & \frac{Q^{2}}{2\, P\cdot q}=const,\qquad t\equiv\Delta^{2}=\left(p'-p\right)^{2}\ll Q^{2},\end{eqnarray}
 where $q$ is the momentum of the virtual photon, $p$ is the momentum
of the inital hadron, $p'$ is the momentum of the final hadron, and
$t$ is the momentum transfer. In this kinematics the DVCS amplitude
is factorized~\cite{Ji:1998xh,Collins:1998be} into the convolution
of a perturbative coefficient function with a soft matrix element--generalized
parton distribution (GPD) of the parton inside the target.

While for the region $x\sim1$ the dominant contribution to the DVCS
amplitude comes from the generalized parton distributions (GPD) for
quarks, we know from experience with deep-inelastic scattering (DIS)
that in the small-$x$ kinematics (large c.m. energy $W$) the density
of  partons (especially gluons) increases and the dominant contribution
comes from the gluon sector.

In this paper we employ the color dipole approch introduced in~\cite{Kopeliovich:1981,mueller}.
The central object of the model is the dipole scattering amplitude,~$\mathcal{A}(x,\vec{r})$.
For DIS, only the imaginary part of the amplitude contributes, and
the result becomes especially simple~\cite{Kopeliovich:1981,Nikolaev:1994uu},
\begin{equation}
\sigma_{\gamma^{*}p}\left(x,Q^{2}\right)=\int_{0}^{1}dz\int d^{2}r\,\left|\Psi\left(z,r,Q^{2}\right)\right|^{2}\sigma_{\bar{q}q}(x,r),\label{eq:XSection-DIS}\end{equation}

-the DIS cross-section equals the dipole cross-section~$\sigma_{\bar{q}q}(x,r)$
 ``averaged'' with weight $\left|\Psi\left(z,r,Q^{2}\right)\right|^{2}$,
where $\Psi\left(z,r,Q^{2}\right)$ is the virtual photon light-cone
wave function, which depends on transverse separation $\vec{r}$ and
fraction $z$ of the light-cone momentum carried by the quark (see
Section~\ref{sec:WFfromIVM} for rigorous definition).

Generalization of~(\ref{eq:XSection-DIS}) to the DVCS case is straightforward
\cite{McDermott:2001pt,Favart:2003cu},
although one has to pay attention to some subtle points. First,  one has to deal with the nonperturbative contribution of large-size
dipoles. While it also exists in the forward case (DIS), the problem
is exacerbated in the off-foward case (DVCS) due to presence of the real
photon in the final state. \emph{A priori}, one cannot say how important
this contribution is in a particular process. As a first approximation,
one can just ignore the nonperturbative effects, as was done in~\cite{McDermott:2001pt,Favart:2003cu,Machado:2007zz,Machado:2008tp},
and evaluate everything perturbatively. While for DIS this approximation
is partially justified due to high virtualities $Q^{2}$ in both vertices, for
DVCS its validity is questionable, since  the final photon is real.
In the present evaluations we rely on the instanton vacuum model for the
photon wave function.

Another complication is that in
contrast to DIS, the DVCS amplitude is sensitive to both the real
and imaginary parts of the dipole amplitude. In this evaluations we
restore the real part using the prescription proposed in~\cite{Bronzan:1974jh}
(see Section~\ref{sec:DipoleModel} for more details).

The paper is organized as follows. In Section~\ref{sec:DipoleModel}
we review the color dipole model and present the parameters used in
numerical evaluations. In Section~\ref{sec:WFfromIVM} we evaluate
the photon wave function within the instanton vacuum model. In Sections~\ref{sec:Results}
and~\ref{sec:Conclusions} we present results for the DVCS cross-sections,
compare with available data, and draw conclusions.

\section{Color dipole model}

\label{sec:DipoleModel} As was mentioned above, the color dipole
model is valid only in the region of small-$x$, where the dominant
contribution to the DVCS amplitude comes from  gluonic exchanges.
The general expression for the DVCS amplitude in the color dipole
model has the form,

\begin{equation}
\mathcal{A}_{\mu\nu}\left(x_{B},t,Q^{2}\right)\propto\epsilon_{\mu}^{(i)}\epsilon_{\nu}^{(j)}\int d\beta_{1}d\beta_{2}d^{2}r_{1}d^{2}r_{2}\bar{\Psi}^{(i)}\left(\beta_{2},\vec{r}_{2},Q^{2}=0\right)\mathcal{A}_{ij}^{d}\left(\beta_{1},\vec{r}_{1};\beta_{2},\vec{r}_{2};Q^{2},\Delta\right)\Psi^{(j)}\left(\beta_{1},\vec{r}_{1},Q^{2}\right),\label{eq:Convolution:Full}\end{equation}
where $\beta_{1,2}$ are the light-cone fractional momenta of the
quark, and $\vec{r}_{1,2}$ are the transverse distances in the final
and initial states respectively (in off-forward kinematics they are
no longer equal), $\Delta$ is the momentum transfer in DVCS, $\mathcal{A}^{d}(...)$
is the scattering amplitude for the dipole state, and indices $(i,j)$
refer to polarizations of inital and final photons. In general case
the amplitude $\mathcal{A}^{d}(...)$ is a nonperturbative object,
with asymptotic behaviour for small~$r$ determined from the pQCD
\cite{Kopeliovich:1981}:\begin{equation}
\mathcal{A}^{d}(...)\sim r^{2}\end{equation}
 (up to slowly varying corrections $\sim\ln(r)$).

In our approach we use a model~\cite{Kopeliovich:2007fv,Kopeliovich:2008nx,Kopeliovich:2008da}
for the partial amplitude of the scattering of the color dipole on
the nucleon, which has the form

\begin{eqnarray}
\mathcal{A}_{ij}^{d}\left(\beta_{1},\vec{r}_{1};\beta_{2},\vec{r}_{2};Q^{2},\Delta\right) & \approx & \delta\left(\beta_{1}-\beta_{2}\right)\delta\left(\vec{r}_{1}-\vec{r}_{2}\right)\int d^{2}b\, e^{i\vec{\Delta}\vec{b}}Im\, f_{\bar{q}q}^{N}(\vec{r}_{1},\vec{b},\beta_{1})\label{eq:DVCSIm-BK}\\
\vec 
Im\, f_{\bar{q}q}^{N}(\vec{r},\vec{b},\beta) & = & \frac{1}{12\pi}\int\frac{d^{2}k\, d^{2}\Delta}{\left(\vec k+\frac{\vec \Delta}{2}\right)^{2}\left(\vec k-\frac{\vec \Delta}{2}\right)^{2}}\alpha_{s}\mathcal{F}\left(x,\vec{k},\vec{\Delta}\right)\, e^{i\vec b\cdot\vec \Delta}\nonumber\\
 & \times & \left(e^{-i\beta \vec r\cdot\left(\vec k-\frac{\vec \Delta}{2}\right)}-e^{i(1-\beta)\vec r\cdot\left(\vec k-\frac{\vec \Delta}{2}\right)}\right)\left(e^{i\beta \vec r\cdot\left(\vec k+\frac{\vec \Delta}{2}\right)}-e^{-i(1-\beta)\vec r\cdot\left(\vec k+\frac{\vec \Delta}{2}\right)}\right),
 \label{6}
 \end{eqnarray}
where $\vec{b}$ is impact parameter; $\left(\vec{k}-\frac{\vec{\Delta}}{2},\vec{k}+\frac{\vec{\Delta}}{2}\right)$
are the transverse momenta of the incoming and outgoing quarks; $\mathcal{F}\left(x,\vec{k},\vec{\Delta}\right)$
is the \emph{unintegrated} GPD for gluons~
\footnote{The exact definition and number of independent GPDs depends on the
spin. Below we consider only the simplest case of spin-$0$ target.
Modelling of the helicity components is a much more complicated task
which contains large uncertainties.},
\begin{eqnarray}
\frac{\mathcal{F}\left(x,\vec{k},\vec{\Delta}\right)}{k^{2}}&\equiv& H\left(x,\vec{k},\vec{\Delta}\right)\\
&=&\nonumber\frac{1}{2}\int d^{2}r\, e^{ik\cdot r}\int\frac{dz^{-}}{2\pi}e^{ix\bar{P}^{+}z^{-}}\left\langle P'\left|\bar{\psi}\left(-\frac{z}{2},-\frac{\vec{r}}{2}\right)\gamma_{+}\mathcal{L}_{\infty}\left(-\frac{z}{2},-\frac{\vec{r}}{2};\frac{z}{2},\frac{\vec{r}}{2}\right)\psi\left(\frac{z}{2},\frac{\vec{r}}{2}\right)\right|P\right\rangle ,
\end{eqnarray}
where $\mathcal{L}_{\infty}\left(x;y\right)$ is the Wilson factor
required by gauge covariance. For this GPD we use the gaussian parameterization~\cite{Kopeliovich:2007fv,Kopeliovich:2008nx,Kopeliovich:2008da},

\begin{equation}
\mathcal{F}\left(x,\vec{k},\vec{\Delta}\right)=\frac{3\sigma_{0}(x)}{16\pi^{2}\alpha_{s}}\left(\vec k+\frac{\vec \Delta}{2}\right)^{2}\left(\vec k-\frac{\vec \Delta}{2}\right)^{2}R_{0}^{2}(x)\exp\left(-\frac{R_{0}^{2}(x)}{4}\left(\vec{k}^{2}+\frac{\vec{ \Delta}^{2}}{4}\right)\right)\exp\left(-\frac{1}{2}B(x)\vec \Delta^{2}\right),\label{eq:GPDParametrization}\end{equation}
where $\sigma_{0}(x),\, R_{0}^{2}(x),\, B(x)$ are  free parameters
fixed from the DIS and $\pi p$ scattering data. We shall discuss
them in more detail in Section~\ref{sec:Results}. The parameterization~(\ref{eq:GPDParametrization})
does not depend on the longitudinal momentum transfer and decreases
exponentially as a function of $\Delta^{2}$. Since it is an effective parameterization valid only in the small-$x$ region, we
do not assume that it satisfies general requirements, such as positivity~\cite{Pobylitsa:2002gw}
and polynomiality~\cite{Ji:1998pc} constraints.

\begin{figure}
\includegraphics[scale=0.3]{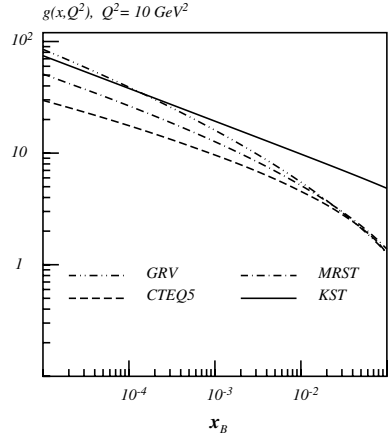}

\caption{\label{fig:BK-CTEQ}Comparison of $k$-integrated forward limit $g_{BK}\left(x,Q^{2}\right)$
of parameterization~(\ref{eq:GPDParametrization}) with realistic
parameterizations: GRV~\cite{Gluck:2007ck}, MRST~\cite{Martin:2006qz}
and CTEQ5~\cite{Lai:1999wy}.}

\end{figure}

The prefactor $\left(\vec k+\frac{\vec \Delta}{2}\right)^{2}\left(\vec k-\frac{\vec \Delta}{2}\right)^{2}$
in~(\ref{eq:GPDParametrization}) guarantees convergence of the integrals
in equation~(\ref{eq:DVCSIm-BK}). In the forward limit,
the amplitude (\ref{eq:DVCSIm-BK}) reduces to the saturated parameterization
of the dipole amplitude proposed by Golec-Biernat and W\"usthoff (GBW)~\cite{GolecBiernat:1998js}.
The corresponding $k$-integrated forward limit of the GPD~(\ref{eq:GPDParametrization}) is compared
with realistic parameterizations of the gluon PDFs - GRV~\cite{Gluck:2007ck},
MRST~\cite{Martin:2006qz} and CTEQ5~\cite{Lai:1999wy} 
in the Figure~(\ref{fig:BK-CTEQ}). We find that in the region $x\lesssim10^{-3}$
the parameterization~(\ref{eq:GPDParametrization}) agrees, within
experimental and theoretical uncertainties, with these realistic parameterizations,
but deviates for larger $x$.

Taking the integrals in~(\ref{eq:DVCSIm-BK}), we arrive at

\begin{eqnarray}
Im\, f_{\bar{q}q}^{N}(\vec{r},\vec{\Delta},\beta) & =\int d^{2}b\, e^{i\vec{b}\vec{\Delta}}Im\, f_{\bar{q}q}^{N}(\vec{r},\vec{b},\beta)=\frac{\sigma_{0}(s)}{4}\exp\left(-\left(\frac{B(s)}{2}+\frac{R_{0}^{2}(s)}{16}\right)\vec{\Delta}_{\perp}^{2}\right)\times\nonumber \\
 & \times\left(e^{-i\beta \vec r\cdot\vec \Delta}+e^{i(1-\beta)\vec r\cdot\vec \Delta}-2e^{i\left(\frac{1}{2}-\beta\right)\vec r\cdot\vec \Delta}e^{-\frac{\vec r^{2}}{R_{0}^{2}(x)}}\right).\label{eq:Im_f-result}\end{eqnarray}

Evaluation of the real part of the amplitude is quite straightforward.
As it was discussed in~\cite{Bronzan:1974jh}, if the limit $\lim_{s\to\infty}\left(\frac{\mathcal{I}m\,\mathcal{A}}{s^{\alpha}}\right)$
exists and is finite, then the real part of the amplitude is related
with imaginary part by\begin{equation}
\mathcal{R}e\,\mathcal{A}=s^{\alpha}\tan\left[\frac{\pi}{2}\left(\alpha-1+\frac{\partial}{\partial\ln s}\right)\right]\frac{\mathcal{I}m\,\mathcal{A}}{s^{\alpha}}.\label{eq:BronzanFul}\end{equation}

As it will be shown in Section~\ref{sec:Results}, the imaginary
part in the color dipole model indeed has a power dependence on energy,
$\mathcal{I}m\,\mathcal{A}(s)\sim s^{\alpha}$. The formula~(\ref{eq:BronzanFul})
in this case may be simplified to \begin{eqnarray}
\frac{\mathcal{R}e\,\mathcal{A}}{\mathcal{I}m\,\mathcal{A}} & =\tan\left(\frac{\pi}{2}(\alpha-1)\right):=\eta.\end{eqnarray}

Then for the DVCS amplitude we finally obtain \begin{equation}
\mathcal{A}_{\mu\nu}^{(ij)}\approx(\eta+i)\epsilon_{\mu}^{(i)}(q')\epsilon_{\nu}^{(j)}(q)\int d^{2}r\int d\beta\bar{\Psi}^{(i)}(\beta,r,Q^{2}=0)\Psi^{(j)}(\beta,r,Q^{2})\, Im\, f_{\bar{q}q}^{N}(\vec{r},\vec{\Delta},\beta),\label{eq:DVCS-Im-conv}\end{equation}
where in~(\ref{eq:DVCS-Im-conv}) the upper indices $(ij)$ refer
to polarizations of the final and initial states, and we evaluated
the real part of the DVCS amplitude according to (\ref{eq:BronzanFul}).
For the cross-section of unpolarised DVCS amplitude, from~(\ref{eq:DVCS-Im-conv})
we obtain\footnote{The DVCS amplitude in color dipole approach is defined as $\mathcal{A}\sim\frac{x}{Q^{2}}\epsilon_{\mu}^{(i)}\epsilon_{\nu}^{(j)}\left\langle p'\left|J_{\mu}(z)J_{\nu}(0)\right|p\right\rangle ,$
the extra kinematic prefactor $x^{2}/Q^{4}$ is absorbed in the amplitude.}

\begin{eqnarray}
\frac{d\sigma}{dt} & = & \frac{1+\eta^{2}}{16\pi}\sum_{ij}\left|\mathcal{A}_{\mu\nu}^{(ij)}\right|^{2}=\nonumber \\
 & = & \frac{1+\eta^{2}}{16\pi}\sum_{ij}\left|\int d^{2}r\int d\beta\bar{\Psi}^{(i)}(\beta,r,Q^{2}=0)\Psi^{(j)}(\beta,r,Q^{2})\, i\, Im\, f_{\bar{q}q}^{N}(\vec{r},\vec{\Delta},\beta)\right|^{2}.\label{eq:DVCS-cross-section}\end{eqnarray}

Since the helicity-flip scattering is a higher-twist effect, it is
suppressed in the large-$Q^{2}$ kinemtics. Thus we may further simplify~(\ref{eq:DVCS-cross-section})
to

\begin{equation}
\frac{d\sigma}{dt}\approx\frac{1+\eta^{2}}{16\pi}\sum_{i}\left|\int d^{2}r\int d\beta\bar{\Psi}^{(i)}(\beta,r,Q^{2}=0)\Psi^{(i)}(\beta,r,Q^{2})\, i\, Im\, f_{\bar{q}q}^{N}(\vec{r},\vec{\Delta},\beta)\right|^{2}.\label{eq:DVCS-cross-section-largeQ2}\end{equation}

Formula~(\ref{eq:DVCS-cross-section-largeQ2}) is the final result
which will be used for numerical evaluation of the DVCS amplitude.
It is important to note that in the small-$r$ region the amplitude~(\ref{eq:Im_f-result})
behaves as $\mathcal{A}\sim r^{2}$, therefore the corresponding cross-section~(\ref{eq:DVCS-cross-section})
is falling as $1/Q^{4}$ at large-$Q^{2}$, in agreement with the
general analysis~\cite{Mueller:1998fv,Ji:1996nm,Ji:1998xh}.

\section{Photon wave function in the instanton vacuum}

\label{sec:WFfromIVM}In this section we would like to provide some
details of evaluation of the wave function in the instanton vacuum
model (IVM)~(see~\cite{Schafer:1996wv,Diakonov:1985eg,Diakonov:1995qy,Goeke:2007bj,Dorokhov:2006qm}
and references therein). In the leading order in $N_{c}$, the model has the same Feynman rules as
in the perturbative theory, but with momentum-dependent quark mass
$\mu(p)$ in the quark propagator~\cite{Diakonov:1985eg}\begin{eqnarray}
S(p) & = & \frac{1}{\hat{p}-\mu(p)+i0},\end{eqnarray}
 and nonlocal interaction vertex of vector current~\cite{Goeke:2007bj}
\begin{eqnarray}
&&\hat{v}\equiv v_{\mu}\gamma^{\mu}\rightarrow\hat{V}=\hat{v}-M\left(G_{\mu}(p,q)f(p+q)+G_{\mu}(p+q,-q)f(p)\right)v^{\mu}(q),\label{eq:Vertex_nonlocal}\\
&&G_\mu(p,q)\equiv \sum_{n=0}^\infty \frac{1}{(n+1)!}f_{,\mu,\mu_1...\mu_n}q_{\mu_1}...q_{\mu_n}\approx f_{,\mu}(p)+\mathcal{O}(q),
\end{eqnarray}
 where $p,\, p+q$ are the momenta of the incoming and outgoing quarks
respectively, and $f_{,\mu_1...\mu_n}(p)\equiv\partial^n f(p)/\partial p_{\mu_1}...\partial p_{\mu_n}$.
Using symmetry properties of the the last term in (\ref{eq:Vertex_nonlocal}) and properties of the function $G_\mu(p,q)$, it is possible to cast (\ref{eq:Vertex_nonlocal}) to the equivalent form
\begin{equation}
\hat{V}=v^\mu(q)\left(\gamma_\mu-(2p_\mu+q_\mu) \frac{M\left(f^2(p+q)-f^2(p)\right)}{(p+q)^2-p^2}\right),
\end{equation}
which is frequently used~\cite{Anikin:2000rq,Dorokhov:2003kf}.

The mass of the constituent quark has a form~\cite{Diakonov:1985eg} 
\begin{equation}
\mu(p)=m+M\, f^{2}(p),
\end{equation}
 where $m\approx5$~MeV is the current quark mass, and $M\approx350$~MeV
is the contribution of the instanton-induced effects. In the limit
$p\to\infty$ the instanton-induced nonlinear formfactor $f(p)$ falls
off as $\sim\frac{1}{p^{3}},$ so for large $p\gg\rho^{-1}$, where
$\rho\approx(600\, MeV)^{-1}$ is the average instanton size, the
mass of the quark $\mu(p)\approx m$ and the vector current interaction
vertex $\hat{V}\approx\hat{v}$. However, we would like to emphasize
that all the correlators get contributions from both
the soft and the hard parts, so even in the large-$Q$ limit the instanton
vacuum wave function is different from the well-known QED result.

The overlap of the initial and final photon wave functions in~(\ref{eq:DVCS-cross-section-largeQ2})
was evaluated according to
\begin{equation}
\Psi^{(i)*}(\beta,r,Q^{2}=0)\Psi^{(i)}(\beta,r,Q^{2})=\sum_{\Gamma}I_{\Gamma}^{*}\left(\beta,r^{*},0\right)I_{\Gamma}\left(\beta,r,Q^{2}\right),\label{eq:ConvolutionWF}\end{equation}
 where the summation is done over possible polarization states $\Gamma_{em}=\{\gamma_{\mu},\gamma_{\mu}\gamma_{5},\sigma_{\mu\nu}\}$, and $I_\Gamma$ corresponds to one of the matrix elements
\begin{eqnarray}
\label{eq:WFCompDef-Vec}I_{\mu}(\beta,\vec{r}) & = & \int\frac{dz^{-}}{2\pi}e^{i\left(\beta+\frac{1}{2}\right)q^{-}z^{+}}\left\langle 0\left|\bar{\psi}\left(-\frac{z}{2},-\frac{\vec{r}}{2}\right)\gamma_{\mu}\psi\left(\frac{z}{2},\frac{\vec{r}}{2}\right)\right|\gamma(q)\right\rangle ,\\
\label{eq:WFCompDef-Pseudo}I_{\mu}^{5}(\beta,\vec{r}) & = & \int\frac{dz^{-}}{2\pi}e^{i\left(\beta+\frac{1}{2}\right)q^{-}z^{+}}\left\langle 0\left|\bar{\psi}\left(-\frac{z}{2},-\frac{\vec{r}}{2}\right)\gamma_{\mu}\gamma_{5}\psi\left(\frac{z}{2},\frac{\vec{r}}{2}\right)\right|\gamma(q)\right\rangle ,\\
\label{eq:WFCompDef-Tensor}I_{\mu\nu}(\beta,\vec{r}) & = & \int\frac{dz^{-}}{2\pi}e^{i\left(\beta+\frac{1}{2}\right)q^{-}z^{+}}\left\langle 0\left|\bar{\psi}\left(-\frac{z}{2},-\frac{\vec{r}}{2}\right)\sigma_{\mu\nu}\psi\left(\frac{z}{2},\frac{\vec{r}}{2}\right)\right|\gamma(q)\right\rangle .\end{eqnarray}
Notice also that in the final state in~\ref{eq:ConvolutionWF} we should use $r_{\mu}^{*}=r_{\mu}+n_{\mu}\frac{q_{\perp}'\cdot r_{\perp}}{q_{+}}=r_{\mu}-n_{\mu}\frac{\Delta_{\perp}\cdot r_{\perp}}{q_{+}}$,
which takes into account that the final photon has $q'_\perp\not=0$ whereas the components of the wave function~(\ref{eq:WFCompDef-Vec}-\ref{eq:WFCompDef-Tensor}) are defined in the reference frame with $q_{\perp}=0$.

\begin{figure}
\includegraphics{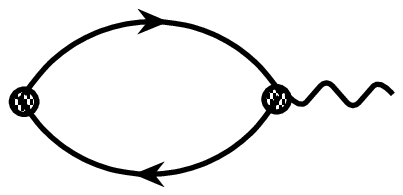}
\caption{\label{fig:WF-LO}In the leading order there is only one diagram which
contributes to the wave function~(\ref{eq:WF-LO}).}
\end{figure}

The wave functions corresponding to matrix elements~(\ref{eq:WFCompDef-Vec}-\ref{eq:WFCompDef-Tensor}) were evaluated in~\cite{Dorokhov:2006qm}.
 In the leading order in $N_{c}$ one can easily obtain for the components $I_\Gamma$
\begin{equation}
I_{\Gamma}=\int\frac{d^{4}p}{(2\pi)^{4}}e^{i\vec{p}_{\perp}\vec{r}_{\perp}}\delta\left(p^{+}-\left(\beta+\frac{1}{2}\right)q^{+}\right)Tr\left(S(p)\hat{V}S(p+q)\Gamma_{em}\right),\label{eq:WF-LO}\end{equation}
 which corresponds to the diagram shown in the Figure~(\ref{fig:WF-LO}).
 The evaluation of (\ref{eq:WF-LO}) is quite straightforward and in the reference frame with $q_\perp=0,\, \epsilon_{\lambda}^{(i)}(q)  =\epsilon_{\lambda_{\perp}}^{(i)}(q)$ yields

\begin{eqnarray}
&&I_{\mu}\left(\beta,r\right)  =\int dz^{-}e^{-i(\beta-1/2)q^{+}z^{-}}\left\langle 0\left|\bar{\psi}\left(-\frac{z}{2},-\frac{\vec{r}}{2}\right)\gamma_{\mu}\psi\left(\frac{z}{2},\frac{\vec{r}}{2}\right)\right|\gamma^{\lambda}(q)\right\rangle \label{eq:WFVecResult}\\
&& =-ie_{q}\epsilon_{\nu}^{(\lambda)}(q)\int\frac{d^{3}p}{(2\pi)^{3}}e^{-ir_{\perp}\left(p+\frac{q}{2}\right)}\times\nonumber \\
&& \times\left(\frac{p_{\mu}(p+q)^{\nu}+p^{\nu}(p+q)_{\mu}-\left(p^{2}+p\cdot q-\mu(p)\mu(p+q)\right)\delta_{\mu}^{\nu}}{\left(2\beta q^{+}p^{-}-p_{\perp}^{2}-\mu^{2}\left(p\right)+i0\right)\left(2\left(1-\beta\right)q^{+}\left(p^{-}-\frac{Q^{2}}{2q^{+}}\right)-p_{\perp}^{2}-\mu^{2}\left(p+q\right)+i0\right)}\right.\nonumber \\
&& -\left.\frac{M\left(G_{\nu}\left(p,q\right)f\left(p+q\right)+f\left(p\right)G_{\nu}\left(p+q,-q\right)\right)\left(\mu(p)(p+q)_{\mu}+\mu(p+q)p_{\mu}\right)}{\left(2\beta q^{+}p^{-}-p_{\perp}^{2}-\mu^{2}\left(p\right)+i0\right)\left(2\left(1-\beta\right)q^{+}\left(p^{-}-\frac{Q^{2}}{2q^{+}}\right)-p_{\perp}^{2}-\mu^{2}\left(p+q\right)+i0\right)}\right),\nonumber\\
&&I_{\mu}^{5}\left(\beta,r\right) = \int dz^{-}e^{-i(\beta-1/2)q^{+}z^{-}}\left\langle 0\left|\bar{\psi}\left(-\frac{z}{2},-\frac{\vec{r}}{2}\right)\gamma_{\mu}\gamma_{5}\psi\left(\frac{z}{2},\frac{\vec{r}}{2}\right)\right|\gamma^{\lambda}(q)\right\rangle \label{eq:I:3Dim5Nonl}\\
&&= -ie_{q}\epsilon_{\mu\alpha\beta\gamma}\epsilon^{\alpha(\lambda)}(q)q^{\beta}4N_{c}\int\frac{d^{3}p}{(2\pi)^{3}}e^{-ir_{\perp}\left(p+\frac{q}{2}\right)}\times\nonumber \\
&& \times  \frac{p_{\gamma}}{\left(2\beta q_{+}p_{-}-p_{\perp}^{2}-\mu^{2}(p)+i0\right)\left(2\left(1-\beta\right)q_{+}\left(p_{-}-\frac{Q^{2}}{2q_{+}}\right)-p_{\perp}^{2}-\mu^{2}(p+q)+i0\right)}\nonumber\\ 
&&I_{\mu\nu}\left(\beta,r\right) = \int dz_{-}e^{-i(\beta-1/2)q_{+}z_{-}}\left\langle 0\left|\bar{\psi}\left(-\frac{z}{2},-\frac{\vec{r}}{2}\right)\sigma_{\mu\nu}\psi\left(\frac{z}{2},\frac{\vec{r}}{2}\right)\right|\gamma^{\lambda}(q)\right\rangle \label{eq:TT:Final-Nonl}\\
&& =  -ie_{q}\epsilon_{\lambda}^{(i)}(q)\int\frac{d^{3}p}{(2\pi)^{3}}e^{-ir_{\perp}\left(p+\frac{q}{2}\right)}\times\nonumber \\
&& \times  \frac{\mu(p)\left(q_{\nu}g_{\mu\lambda}-q_{\mu}g_{\nu\lambda}\right)+\left(\mu(p+q)-\mu(p)\right)\left(p_{\mu}g_{\nu\lambda}-p_{\nu}g_{\mu\lambda}\right)+M\left(G_{\lambda}\left(p,q\right)f\left(p+q\right)+f\left(p\right)G_{\lambda}\left(p+q,-q\right)\right)\left(p_{\mu}q_{\nu}-p_{\nu}q_{\mu}\right)}{\left(2\beta\, q^{+}p^{-}-p_{\perp}^{2}-\mu^{2}\left(p\right)+i0\right)\left(2\left(1-\beta\right)q^{+}\left(p^{-}-\frac{Q^{2}}{2q^{+}}\right)-p_{\perp}^{2}-\mu^{2}\left(p+q\right)+i0\right)}.\nonumber \end{eqnarray}

We evaluated (\ref{eq:WFVecResult}-\ref{eq:TT:Final-Nonl}) numerically.

\section{Results}

\label{sec:Results}

Here we present the results of the evaluation of the DVCS differential cross section with different models for the partial dipole amplitude $f_{\bar qq}^N(\vec r,\vec b)$, and we also use a  photon wave function
calculated either perturbatively, or within the instanton vacuum model.
First we test the amplitude based on the GBW model.   We also perform calculations with the parameterization proposed by Kowalski and Teaney
(KT) for the saturated dipole cross section \cite{Kowalski:2003hm}. In addition, we try an energy dependent KST parameterization  for
the dipole cross section, proposed in~\cite{Kopeliovich:2008nx,Kopeliovich:2007pq,Kopeliovich:2008da}.

\subsection{GBW based partial dipole amplitude}

\label{sub:GBW} First of all, we made evaluations in the
GBW model \cite{GolecBiernat:1998js} extended to the $b$-dependent partial amplitude $f_{\bar qq}^N(\vec r,\vec b)$. This parameterization is fitted
to DIS data at large $Q^{2}$ and small Bjorken $x$. The parameters
in Eq.~(\ref{eq:GPDParametrization}) read, $\sigma_{0}(x)=23.03$
mb=const, $R_{0}(x)=0.4\times(x/x_{0})^{0.144}$fm. where $x_{0}=3.04\times10^{-4}$.
The parameter $B(x)$ in Eq.~(\ref{eq:GPDParametrization}), is related to the $t$-slope of the differential cross section of highly virtual photoproduction of vector mesons \cite{Kopeliovich:2007fv,Kopeliovich:2008nx,Kopeliovich:2008da},
\begin{equation}
B(x)=  B_{\gamma^*p\to\rho p}  -{1\over8}\,R_0^2(x)
\label{slope-x}
\end{equation}
We use the experimental value of the slope $B_{\gamma^*p\to\rho p}(x,Q^2\gg1 GeV^2)\approx 5\,GeV^{-2}$ \cite{zeus}.

\begin{figure}[H]
\includegraphics[scale=0.4]{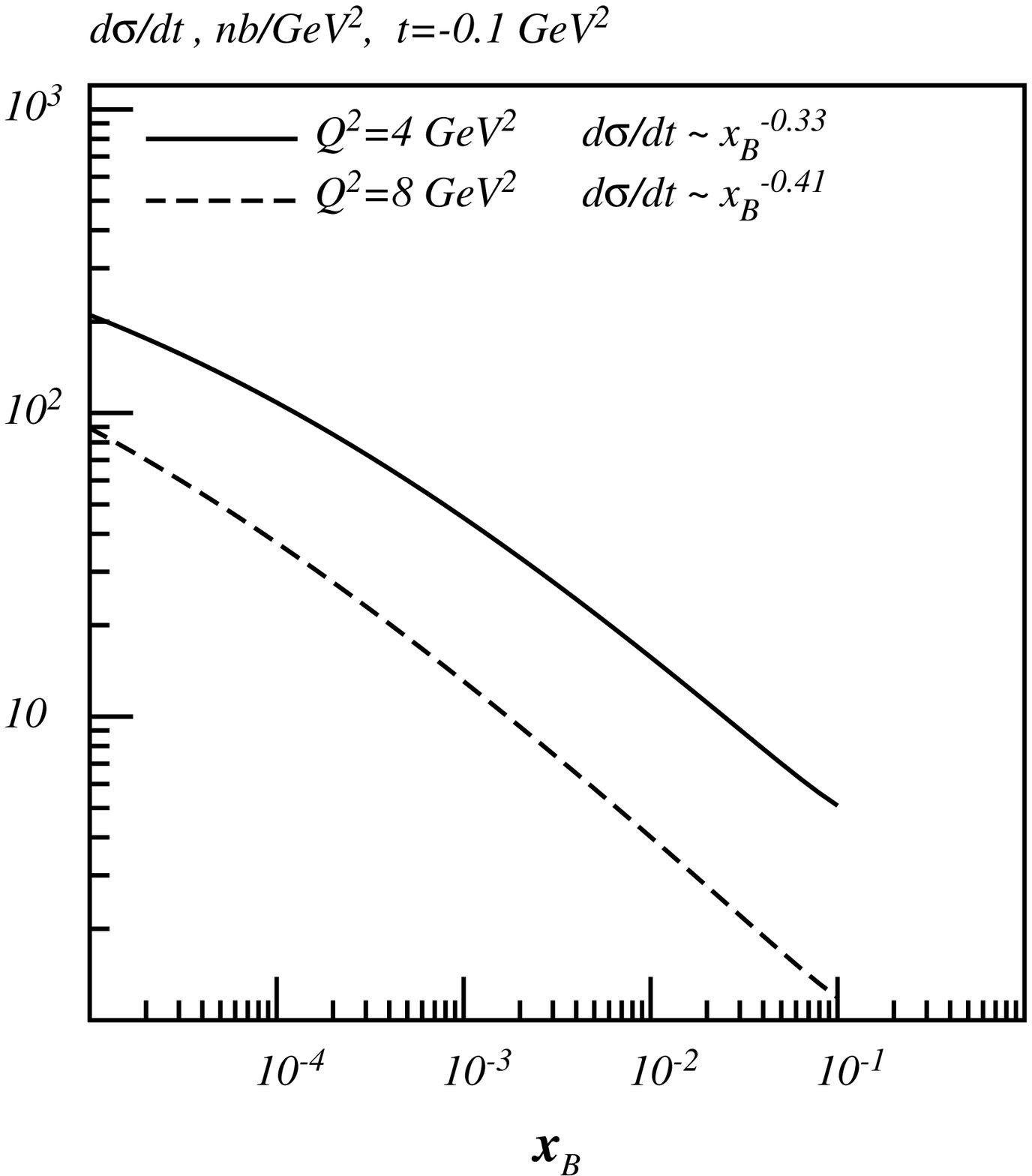}
\includegraphics[scale=0.4]{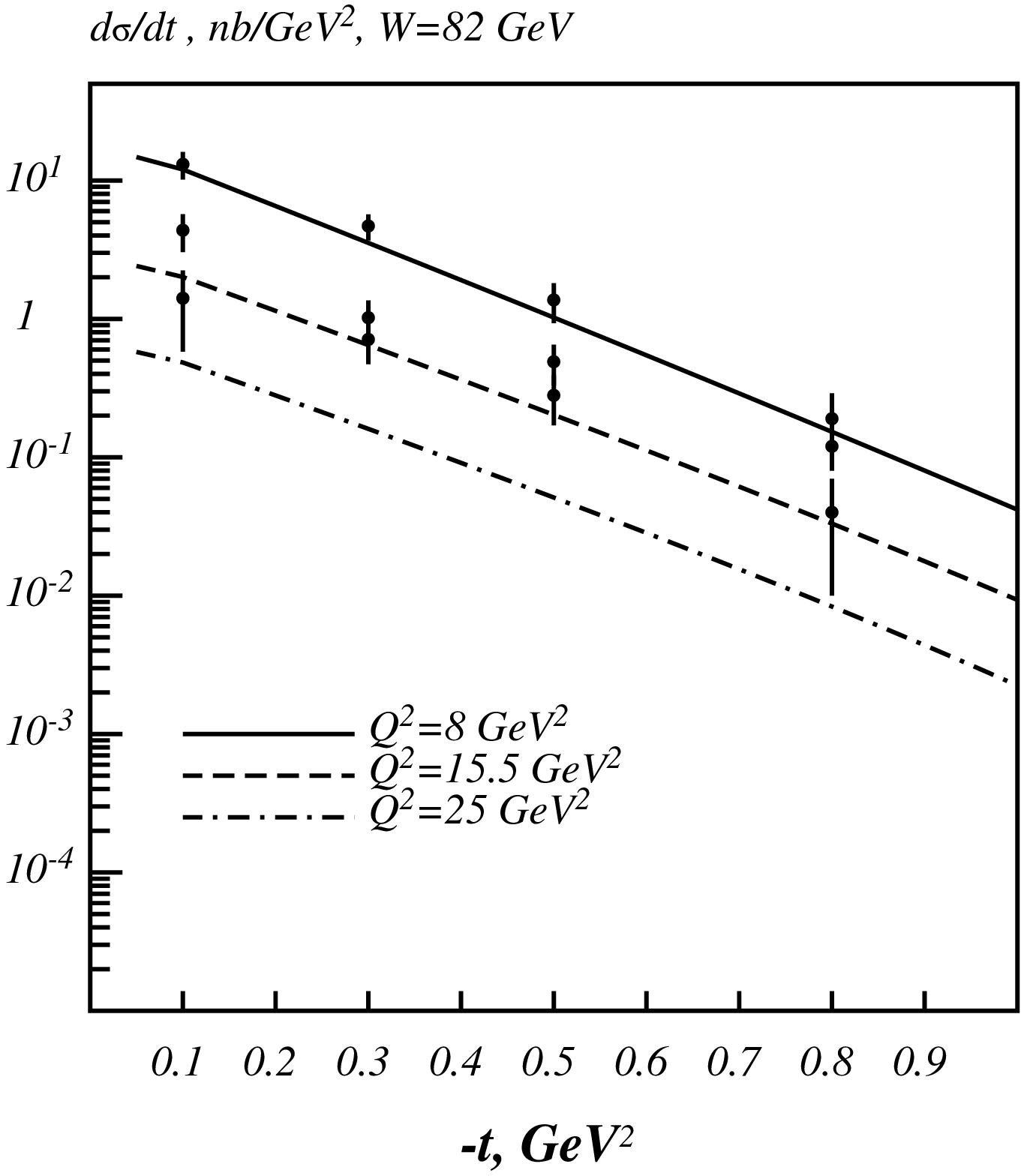}

\caption{\label{fig:DVCS_GBW}$x_{B}-$ and $t-$dependence of the DVCS cross-section
in GBW parameterization}

\end{figure}

Following \cite{GolecBiernat:1998js}, we use the perturbative wave
function of the photon with constituent quark mass $140$\,MeV. From
the left panel of Figure~\ref{fig:DVCS_GBW} we see that in the small-$x_{B}$
region the cross-section is proportional to a power of $x_{B}$, $d\sigma/dt\sim x_{B}^{\alpha}$,
where the power $\alpha$ was obtained by fitting the $x_{B}-$dependence
in the range $x_{B}\in\left(10^{-5},10^{-3}\right)$ and slowly depends
on $\left(Q^{2},\, t\right)$.

In the right panel of Figure~\ref{fig:DVCS_GBW} we compare the model
with data from the H1 experiment at HERA\,\cite{Aaron:2007cz}.
Although the model gives a reasonable description of the data at moderate
values of $Q^{2}$, the discrepancy increases at higher $Q^{2}$.
As it was discussed in Section~\ref{sec:DipoleModel}, the modeled
DVCS cross-section falls as $1/Q^{4}$, while the data behave approximately
as $1/Q^{3}$.

\begin{figure}[H]
\includegraphics[scale=0.4]{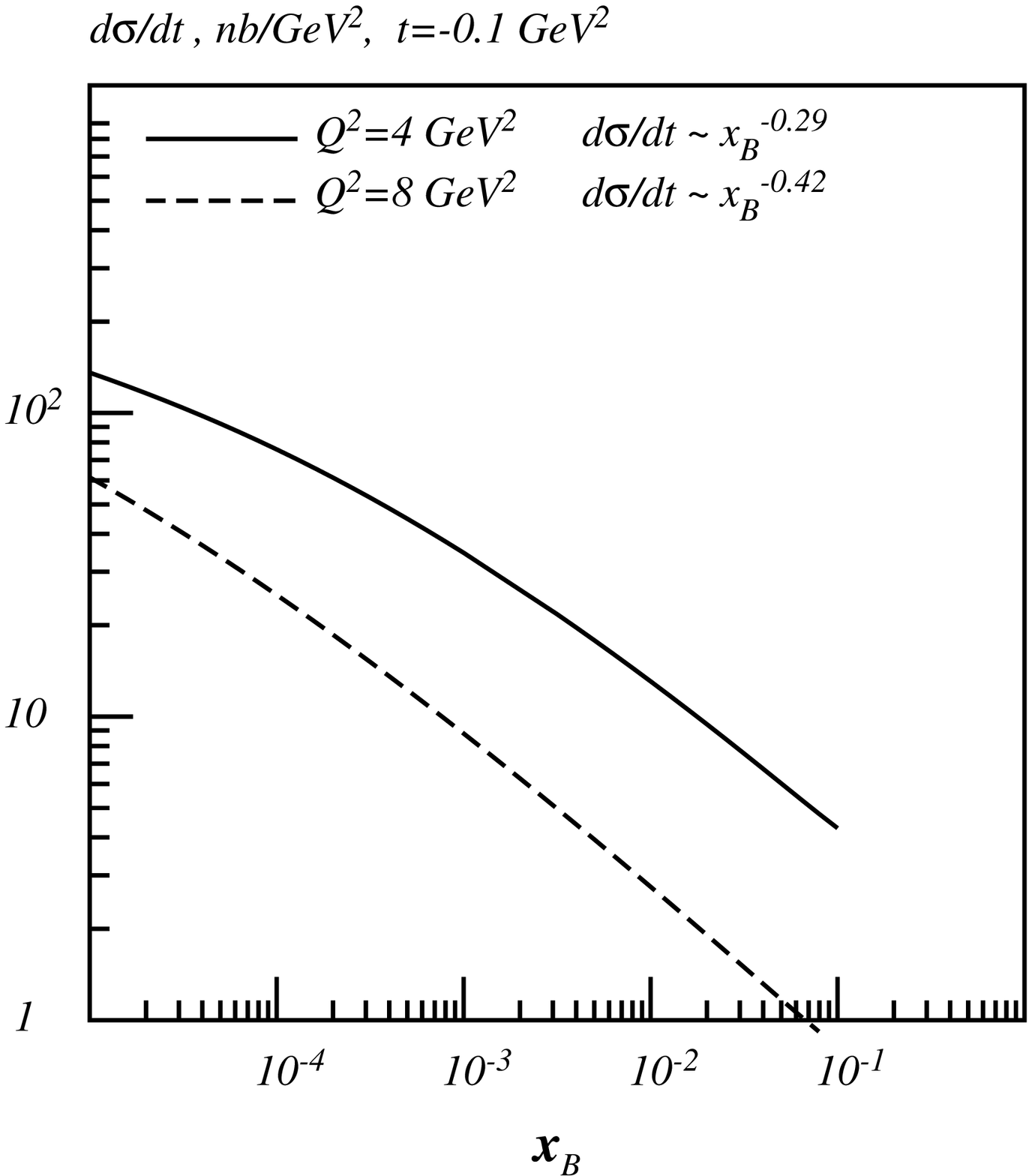}
\includegraphics[scale=0.4]{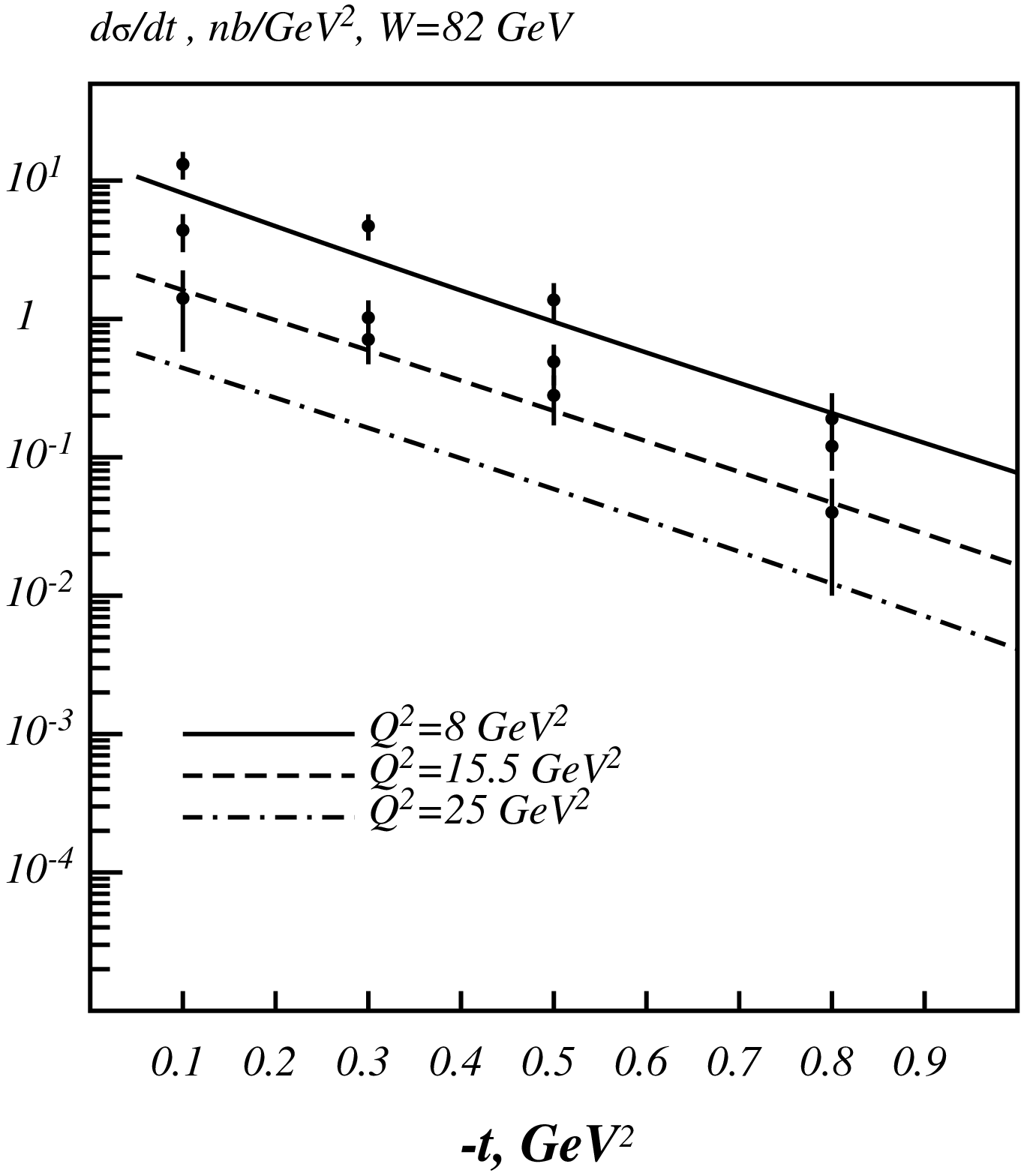}

\caption{\label{fig:DVCS_GBW_IVM}$x_{B}-$ and $t$-dependence of the DVCS
cross-section in GBW parameterization, with realistic photon wave function.}

\end{figure}

In order to identify the source of the disagreement, we replace the
perturbative photon wave function by the one calculated in Section~\ref{sec:WFfromIVM}
within the model of instanton vacuum. The results are depicted in
Figure~\ref{fig:DVCS_GBW_IVM}. We see that the cross sections do
not change very much compared to the previous calculation. This model
gives a reasonable description of the cross-section for moderate $Q^{2}$,
however grossly underestimates the data at high $Q^{2}$.

While the GBW dipole cross section does a pretty good job describing
DIS data, and electroproduction of vector mesons \cite{Hufner:2000jb,Kopeliovich:2001xj},
it fails to explain the observed $Q^{2}$ dependence in DVCS. This
fact shows that DVCS provides a rather sensitive test for models.

\subsection{KT parameterization}

\label{sub:bSat}Another form of impact parameter dependent partial
dipole amplitude, which has correct behavior at small $r$ and the
saturated shape, was proposed in \cite{Kowalski:2003hm}, 
\begin{equation}
{\rm Im}f(\vec{r},\vec{b})=
2\left(1-\exp\left(-\frac{\pi^{2}}{6}r^{2}\alpha_{s}(\mu^{2})\, 
x{g}\left(x,\mu^{2}\right)T_{N}(b)\right)\right),
\label{eq:KT_definition}
\end{equation}
 where the scale $\mu^{2}=0.77\, GeV^{2}+4/r^{2}$; 
 and the gluon distribution function $g\left(x,\mu^{2}\right)$
 was fitted to DIS data. The nucleon profile function in~\cite{Kowalski:2003hm},
$T_{N}(b)$, has a simple form \begin{equation}
T_{N}(b)=\frac{1}{2\pi B_{g}}e^{-b^{2}/2B_{g}},\end{equation} where the slope parameters $B_g=4\,GeV^{-2}$
was fitted to data on electroproduction of vector mesons.

In the small-$r^{2}$ limit this function corresponds to ordinary
gluon PDF $g\left(x,\mu^{2}\right)$.  However, the latter may be different from the results of DGLAP analyses of data, since Eq.~(\ref{eq:KT_definition}) is supposed to include the saturation effects. 
 On the other hand, the eikonalization used in (\ref{eq:KT_definition}) is quite a rough procedure at large $r$ where saturation is at work. A more accurate formula should include a convolution of the dipole amplitude with $T_N(b)$, rather than the simple product. This is why Eq.~(\ref{eq:KT_definition}) misses correlations between $\vec b$ and $\vec r$, which are present in (\ref{6}).

\begin{figure}[H]
\includegraphics[scale=0.4]{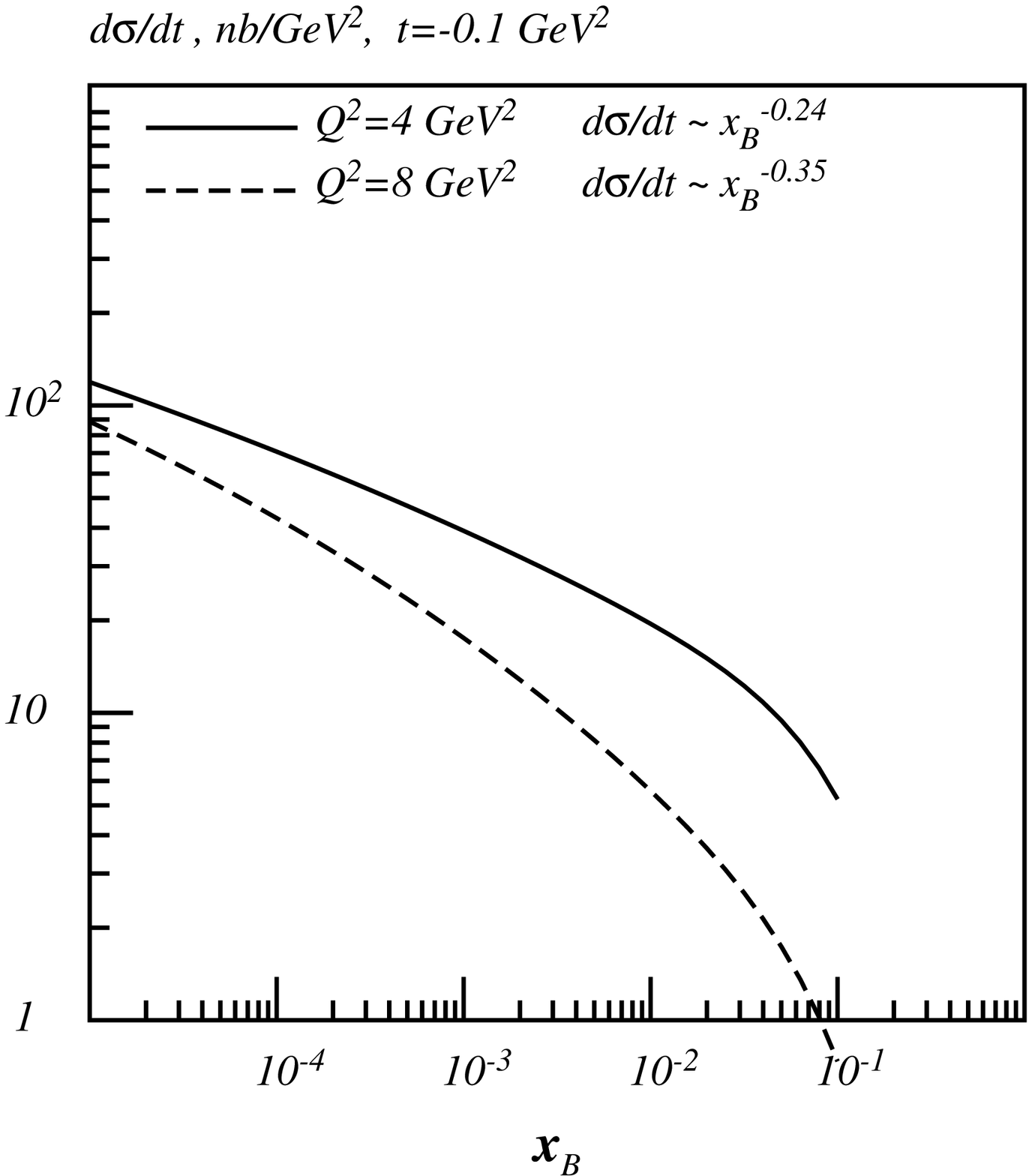}
\includegraphics[scale=0.4]{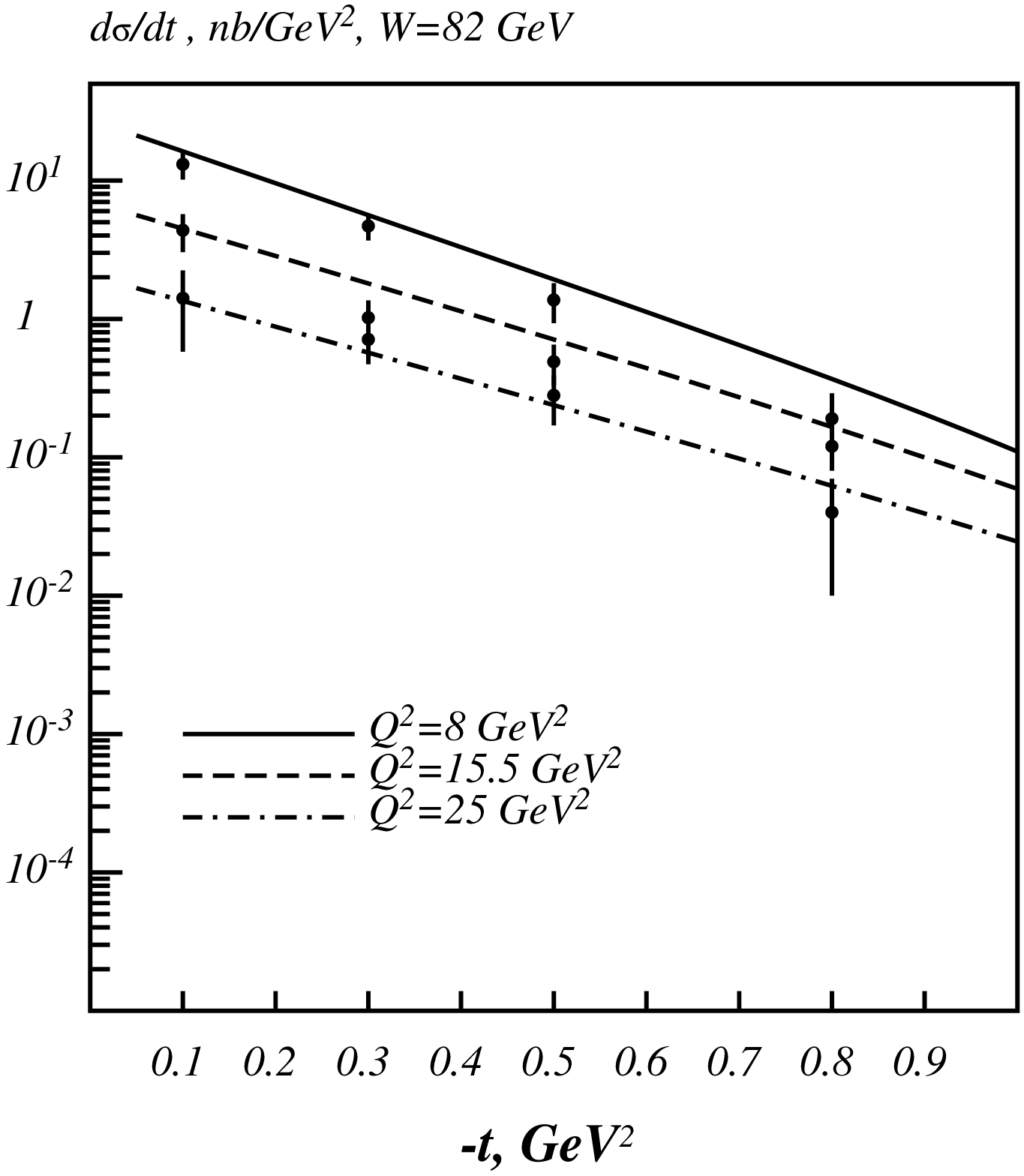}

\caption{\label{fig:DVCS-KT}$x_{B}-$ and $t$-dependence of the DVCS cross-section
in KT parameterization~\cite{Kowalski:2003hm} with the realistic photon
wave function.}

\end{figure}

Results for the DVCS differential cross section, calculated with this
parameterization and the realistic nonperturbative photon wave function,
are presented in Figure~\ref{fig:DVCS-KT}.

\subsection{Energy dependent (KST) parameterization}

\label{sub:KST}

While in the large-$Q^{2}$ kinematics scaling implies that the dipole
amplitude is a function of Bjorken $x$, for smaller $Q^{2}$ (soft
photons) scaling does not work, and for real photons Bjorken $x$
is not an appropriate variable. For this kinematics the photon energy $s$
should be used instead of $x_{B}$. An $s$-dependent parameterization
with saturated form analogous to the GBW was proposed in~\cite{Kopeliovich:1999am},
for the description of the real photo production and absorption, and for DIS
at small $Q^{2}$. One should replace all the $x$-dependent functions
in Eq.~(\ref{eq:GPDParametrization}) by $s$-dependent ones, with

\begin{equation}
\sigma_{0}(s)=\sigma_{tot}^{\pi p}(s)\left(1+\frac{3}{8}\frac{R_{0}^{2}(s)}{\left\langle r_{ch}^{2}\right\rangle }\right),\label{eq:sigma0_paramet}\end{equation}
 where $\sigma_{tot}^{\pi p}(s)=\left(\Sigma_{0}+\Sigma_{1}\ln^{2}(s/s_{0})\right)$
with $\Sigma_{0}=20.9\, mb,\ \Sigma_{1}=0.31\, mb$, and $s_{0}=28.9\, GeV^{2}$,
is the total pion-proton cross section, and $\left\langle r_{ch}^{2}\right\rangle \approx0.44\, fm^{2}$
is the pion charge radius. Correspondingly, $R_{0}(s)=0.88(s/s_{1})^{0.14}\, fm$;
$s_{1}=1000\, GeV^{2}$. In this case the parameter $B(s)$ is related to the $t$-slope of elastic $\pi p$ scattering \cite{Kopeliovich:2008da,Kopeliovich:2008nx},
\begin{equation}
B(s)=B_{el}^{\pi p}(s)-{1\over8}R_0^2(s)-
{1\over3}\left\langle r_{ch}^{2}\right\rangle.
\label{slope-s}
\end{equation}
Naturally, we use the nonperturbative photon
wave function Eq.~(\ref{eq:ConvolutionWF}).

\begin{figure}[H]
\includegraphics[scale=0.4]{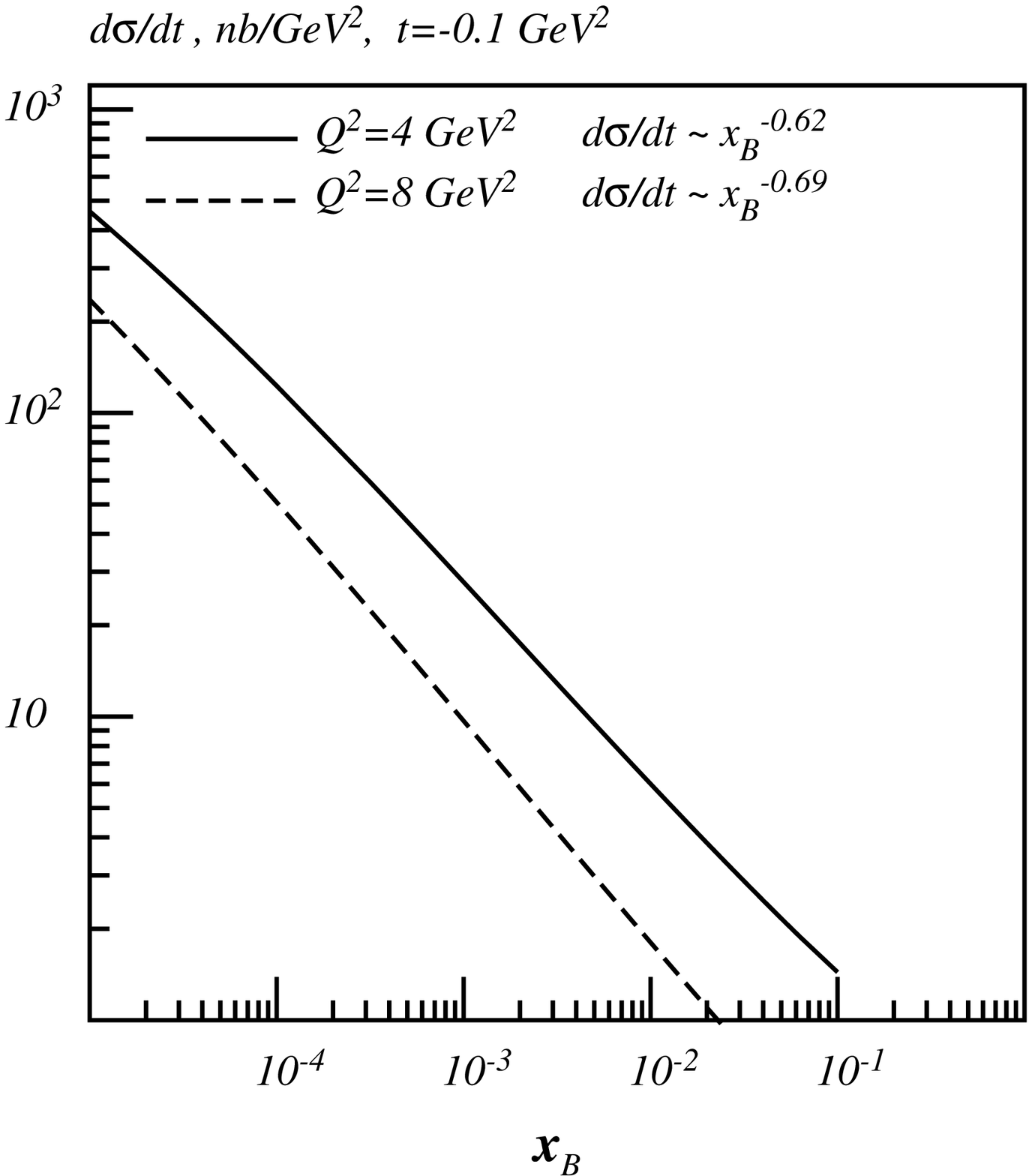}
\includegraphics[scale=0.4]{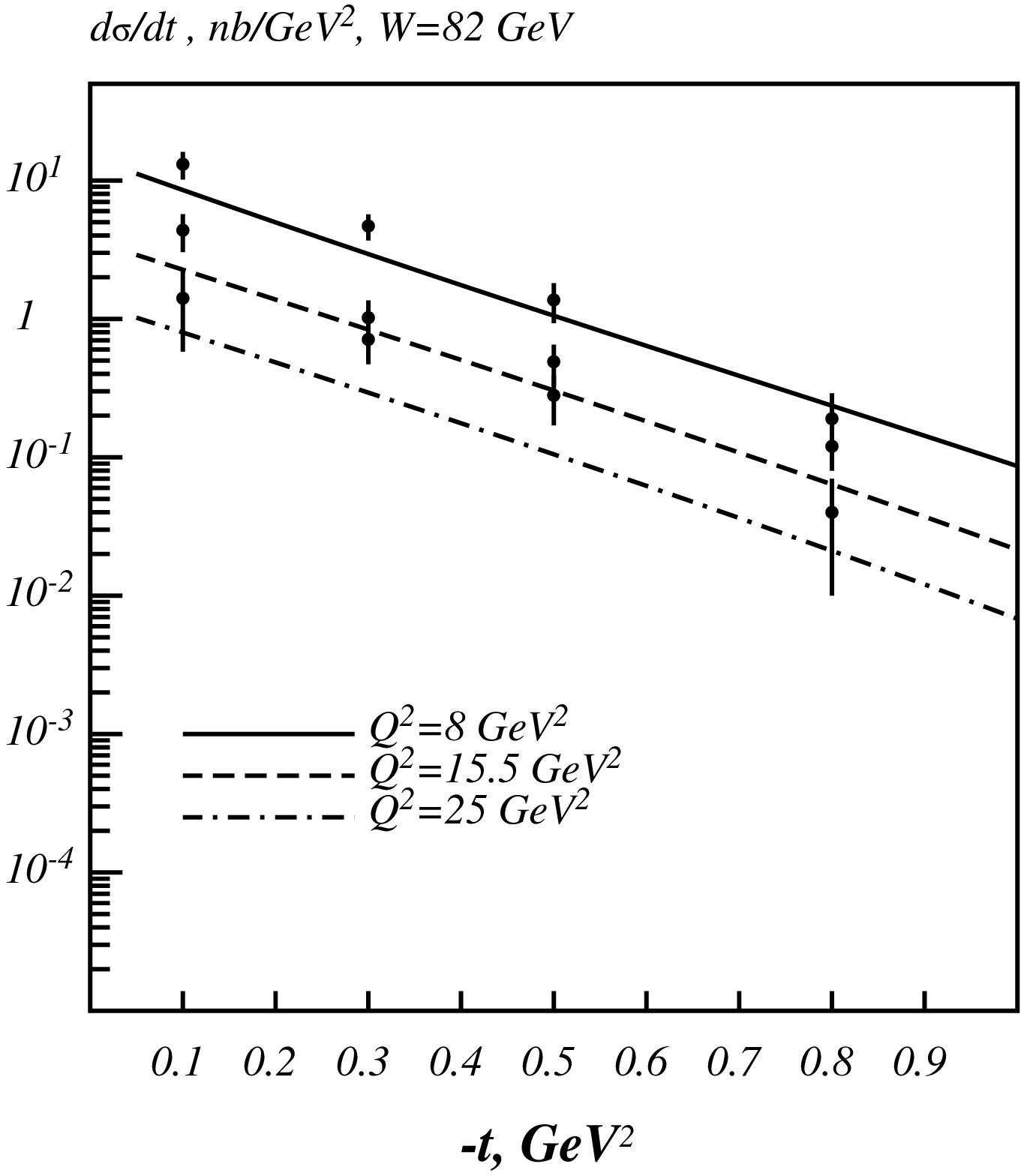}

\caption{\label{fig:DVCS_IVM}$x_{B}-$ and $t$-dependence of the DVCS cross-section,
in KST parameterization, with realistic photon wave function.}

\end{figure}

The numerical results for the DVCS cross section are depicted in Figure~\ref{fig:DVCS_IVM}.
Comparison with data plotted in the right panel show that this model
also leads to a too steep $Q^{2}$ dependence of the cross section,
although its absolute value at $Q^{2}=8\, GeV^{2}$ agrees with the
data. 

\section{Discussion and conclusions}

\label{sec:Conclusions}
\begin{figure}[h]
\includegraphics[scale=0.4]{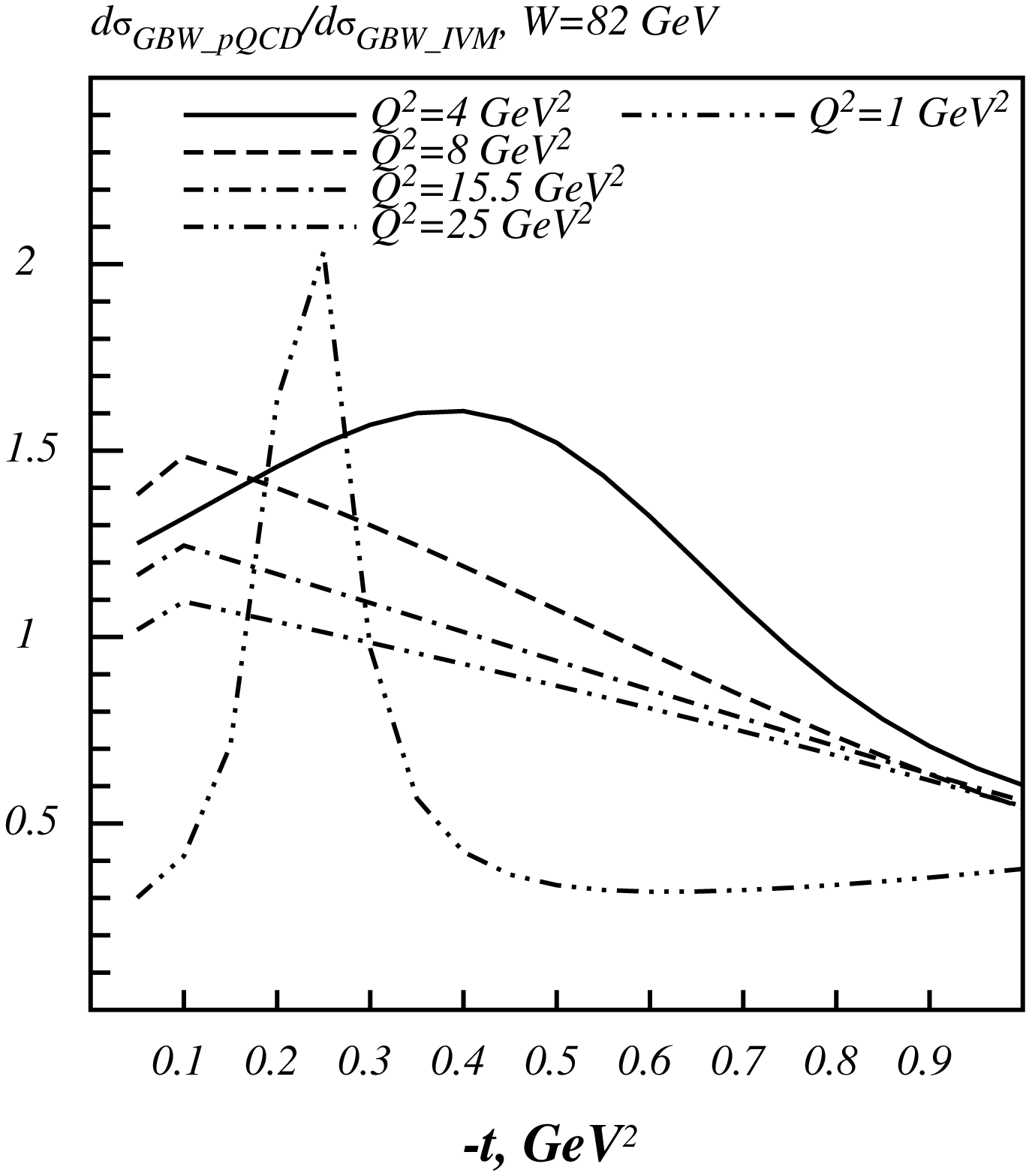}
\includegraphics[scale=0.4]{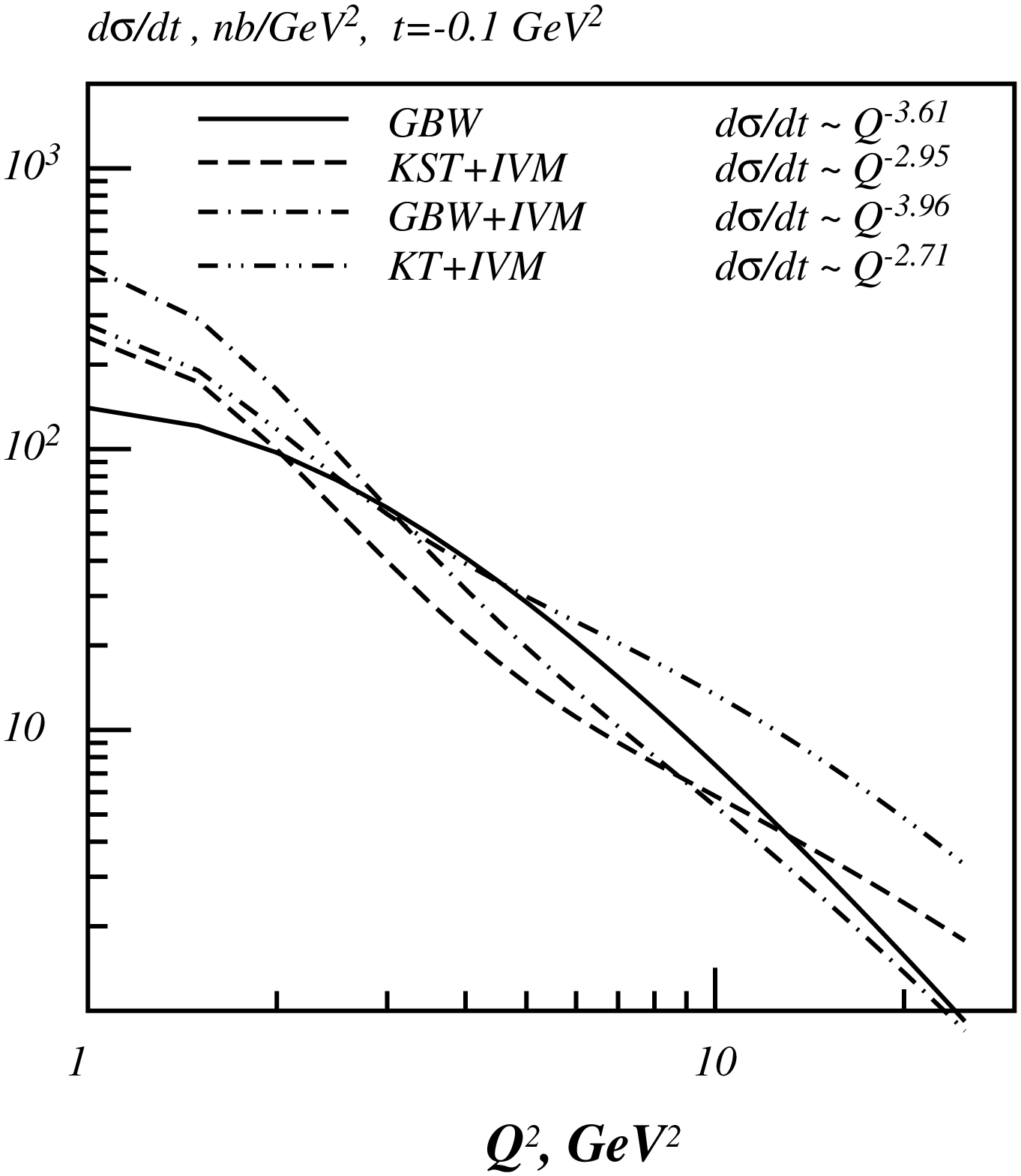}

\caption{\label{fig:bSat-Q2}Left: Ratio of the DVCS cross-section
with pQCD and IVM wave functions  in GBW model. One can clearly see that for $Q^{2}\sim8\, GeV^{2}$
the difference might reach up to 50\other models (GBW and KST). }

\end{figure}

In this paper we evaluated the DVCS cross-section relying on a nonperturbative wave function of the real photon calculated in the instanton vacuum model.
The ratio of DVCS cross sections calculated with the IVM and perturbative wave functions is
shown in the left panel of Figure~\ref{fig:bSat-Q2} as function of $t$ for different values of $Q^2$.

This comparison demonstrates a considerable modification of the cross section 
up to factor two, dependent on kinematics. For small-$Q^2\sim 1$ GeV$^2$ one should use instanton wave functions both for the initial and final photons.
Note that even for $Q^2\sim 1$ GeV$^2$ both cross-sections $d\sigma_{IVM}/dt$ and $d\sigma_{pQCD}/dt$ decrease quite fast as a function of $t$, approximately as $e^{Bt}$, as one can see from the previous Figures~\ref{fig:DVCS_GBW}-\ref{fig:DVCS-KT}. The difference is due to the higher-twist effects which are amplified in the small-$Q^2$ region and become pronounced in the ratio of the cross-sections.

We also tested several models for the dipole cross section. Although all the parameterizations
under discussion have been fitted to DIS data from HERA, not all of them are successful in describing the DVCS cross section, especially the observed $Q^2$ dependence. In particular the dipole partial amplitude based on the popular GBW parameterization of the total cross section leads to a too steep $Q^2$ dependence of the DVCS cross section, calculated with both IVM and perturbative photon wave functions. In this model the cross-section
decreases in the measured interval $Q^2=8-25\,GeV^2$ as $1/Q^{4}$, like one should expect at very large $Q^2$ in accordance with the general
large-$Q^{2}$ analysis~\cite{Mueller:1998fv,Ji:1996nm,Ji:1998xh},
whereas H1 data exhibit an approximate $\sim1/Q^{3}$-behaviour. 
We compare the $Q^{2}$-dependences calculated with different models for the photon wave function and dipole partial amplitude, 
at fixed $x_{B}$ and $t$ in the right panel of Figure~\ref{fig:bSat-Q2}.
We also show in the Figure the results of the fit to the calculated cross sections within the $Q^2$ interval measured in the H1 experiment. 
We conclude that although the absolute value of the cross-section is sensitive
to the nonperturbative effects (large-size dipoles), the $Q^2$-dependence
does not vary much.

Apparently, one can achieve a weaker $Q^2$ dependence by replacing the standard quadratic $r^2$ behavior of the dipole cross section predicted by pQCD at small $r$ \cite{Kopeliovich:1981} by a smaller power of $r$. Such a model \cite{Iancu:2003ge} was considered in \cite{Machado:2008tp}
to describe the $Q^{2}$-dependence of H1 data for DVCS cross section by modifying the small-$r$ behaviour to $\mathcal{A}_{d}\sim r^{1.35+const\ln r}$. This
explains why the model gives reasonable description for the $Q^{2}$-dependence
of the cross-section. 

Remarkably, the KT model \cite{Kowalski:2006hc},
considered above in section~\ref{sub:bSat} provides a reasonable description of the measured $Q^2$, although it has the same
small-$r$ behaviour $\mathcal{A}_{d}\sim r^{2}$ as GBW. This happens because the
higher-twist effects in this model are more pronounced than in GBW.
Straightforward expansion of the dipole amplitude $\mathcal{A}_{d}$
in the KT parameterization yields
(we consider the case $\Delta_{\perp}=0$ for the sake of simplicity),
\begin{eqnarray}
\mathcal{A}_{d} & \sim r^{2}\left[1-\lambda_{KT}(r)\, r^{2}+\mathcal{O}\left(r^{4}\right]\right),\end{eqnarray}
 where the constant $\lambda_{KT}(r)$ depends on $r$ only logarithmically,
and for typical $\langle r\rangle\sim\langle Q\rangle^{-1}\sim0.1\, GeV^{-1}$
we have $\lambda_{KT}(r)\sim0.159\, GeV^{2},$ which is larger than
in GBW.

We would like to emphasize that although it is possible to fit the DIS
data with a parameterization containing sufficient number of free
parameters, not all of the available parameterizations are able to
describe the DVCS data, and a systematic study of the higher-twist
corrections is necessary. In this paper we addressed one of the possible
sources-the higher-twist corrections to the wave function of the initial
photon and considered realistic wave function for the final (real)
photon. Other sources of higher-twist corrections include contributions
of higher-twist quark-gluon operators, such as $\left\langle p'\left|\bar{q}q\right|p\right\rangle ,\left\langle p'\left|\bar{q}G_{+}^{\alpha}G_{+\alpha}q\right|p\right\rangle $
etc. Contrary to DIS and heavy meson electroproduction, in most of
other processes the large-$r$ behaviour of the amplitude \emph{is}
important and gives contribution comparable to the small-$r$ region.
That's why in modelling of the large-$r$ behaviour of the amplitude
we should refer to some microscopic model rather than to a semiphenomenological
approach.

\section*{Acknowledgments}
We would like to thank A. Dorokhov and M. Musakhanov for discussion of the wave function in the instanton vacuum.
This work was supported in part by Fondecyt (Chile) grant 1050589,
and by DFG (Germany) grant PI182/3-1.


\begin{thebibliography}{10}
 \bibitem{Mueller:1998fv} D.~Mueller, D.~Robaschik, B.~Geyer, F.~M.~Dittes
  and J.~Horejsi, Fortsch.\ Phys.\ \textbf{42}, 101 (1994) [arXiv:hep-ph/9812448].
  
 \bibitem{Ji:1996nm} X.~D.~Ji, Phys.\ Rev.\ D \textbf{55}, 7114
  (1997).
  
 \bibitem{Ji:1998pc} X.~D.~Ji, J.\ Phys.\ G \textbf{24}, 1181
  (1998) [arXiv:hep-ph/9807358].
  
 \bibitem{Radyushkin:1996nd} A.~V.~Radyushkin, Phys.\ Lett.\ B
  \textbf{380}, 417 (1996) [arXiv:hep-ph/9604317].
  
 \bibitem{Radyushkin:1997ki} A.~V.~Radyushkin, Phys.\ Rev.\ D
  \textbf{56}, 5524 (1997).
  
 \bibitem{Radyushkin:2000uy} A.~V.~Radyushkin, arXiv:hep-ph/0101225.
  
 \bibitem{Ji:1998xh} X.~D.~Ji and J.~Osborne, Phys.\ Rev.\ D
  \textbf{58} (1998) 094018 [arXiv:hep-ph/9801260].
  
 \bibitem{Collins:1998be} J.~C.~Collins and A.~Freund, Phys.\ Rev.\ D
  \textbf{59}, 074009 (1999).
  
 \bibitem{Collins:1996fb} J.~C.~Collins, L.~Frankfurt and M.~Strikman,
  Phys.\ Rev.\ D \textbf{56}, 2982 (1997).
  
 \bibitem{Brodsky:1994kf} S.~J.~Brodsky, L.~Frankfurt, J.~F.~Gunion,
  A.~H.~Mueller and M.~Strikman, Phys.\ Rev.\ D \textbf{50}, 3134
  (1994).
  
 \bibitem{Goeke:2001tz} K.~Goeke, M.~V.~Polyakov and M.~Vanderhaeghen,
  Prog.\ Part.\ Nucl.\ Phys.\ \textbf{47}, 401 (2001) [arXiv:hep-ph/0106012].
  
 \bibitem{Diehl:2000xz} M.~Diehl, T.~Feldmann, R.~Jakob and P.~Kroll,
  Nucl.\ Phys.\ B \textbf{596}, 33 (2001) [Erratum-ibid.\ B \textbf{605},
  647 (2001)] [arXiv:hep-ph/0009255].
  
 \bibitem{Belitsky:2001ns} A.~V.~Belitsky, D.~Mueller and A.~Kirchner,
  Nucl.\ Phys.\ B \textbf{629}, 323 (2002) [arXiv:hep-ph/0112108].
  
 \bibitem{Diehl:2003ny} M.~Diehl, Phys.\ Rept.\ \textbf{388}, 41
  (2003) [arXiv:hep-ph/0307382].
  
 \bibitem{Belitsky:2005qn} A.~V.~Belitsky and A.~V.~Radyushkin,
  Phys.\ Rept.\ \textbf{418}, 1 (2005) [arXiv:hep-ph/0504030].
  
 \bibitem{Kopeliovich:1981}B.~Z.~Kopeliovich, L.~I.~Lapidus and
  A.~B.~Zamolodchikov, JETP Lett. \textbf{33} (1981) 595 [Pisma
  Zh. Eksp. Teor. Fiz. \textbf{33} (1981) 612].
  
 \bibitem{mueller} A.~H.~Mueller,   Nucl.\ Phys.\ B \textbf{335}, 115 (1990); A.~H.~Mueller and B.~Patel,
   Nucl.\ Phys.\ B \textbf{425}, 471 (1994) [arXiv:hep-ph/9403256].
  
 \bibitem{Nikolaev:1994uu}N.~N.~Nikolaev and B.~G.~Zakharov, Phys.~Lett.~B
  \textbf{327} (1994) 149 [arXiv:hep-ph/9402209].
  
 \bibitem{McDermott:2001pt}
    M.~McDermott, R.~Sandapen and G.~Shaw,
    Eur.\ Phys.\ J.\  C {\bf 22}, 655 (2002)
    [arXiv:hep-ph/0107224].
  
 \bibitem{Favart:2003cu} L.~Favart and M.~V.~T.~Machado,   Eur.\ Phys.\ J.\ C \textbf{29}, 365 (2003) [arXiv:hep-ph/0302079].
  
 \bibitem{Machado:2007zz} M.~V.~T.~Machado,   Braz.\ J.\ Phys.\ \textbf{37} (2007) 555.  
  
 \bibitem{Machado:2008tp}M.~V.~T.~Machado, arXiv:0810.3665 [hep-ph].
  
 \bibitem{Bronzan:1974jh}J.~B.~Bronzan, G.~L.~Kane and U.~P.~Sukhatme,
  Phys.~Lett.~B \textbf{49} (1974) 272.
  
 \bibitem{Kopeliovich:2007fv} B.~Z.~Kopeliovich, H.~J.~Pirner,
  A.~H.~Rezaeian and I.~Schmidt,   Phys.\ Rev.\ D \textbf{77} (2008) 034011 [arXiv:0711.3010 [hep-ph]].
  
 \bibitem{Kopeliovich:2008nx}
    B.~Z.~Kopeliovich, A.~H.~Rezaeian and I.~Schmidt,
    arXiv:0809.4327 [hep-ph], to appear in Phys. Rev. D.
  
  
 \bibitem{Kopeliovich:2008da} B.~Z.~Kopeliovich, I.~K.~Potashnikova,
  I.~Schmidt and J.~Soffer,   arXiv:0805.4534 [hep-ph].  
 \bibitem{Pobylitsa:2002gw}P.~V.~Pobylitsa, Phys.~Rev.~D \textbf{65}
  (2002) 114015 [arXiv:hep-ph/0201030].
  
 \bibitem{GolecBiernat:1998js} K.~J.~Golec-Biernat and M.~W\"usthoff,
  Phys.\ Rev.\ D \textbf{59} (1999) 014017 [arXiv:hep-ph/9807513].
  
 \bibitem{Gluck:2007ck}M.~Gluck, P.~Jimenez-Delgado and E.~Reya,
  Eur. Phys. J. C \textbf{53} (2008) 355 [arXiv:0709.0614 [hep-ph]].
  
 \bibitem{Martin:2006qz}A.~D.~Martin, W.~J.~Stirling and R.~S.~Thorne,
  Phys. Lett. B \textbf{636} (2006) 259 [arXiv:hep-ph/0603143].
  
 \bibitem{Lai:1999wy}H.~L.~Lai \emph{et al.} [CTEQ Collaboration],
  Eur. Phys. J. C \textbf{12} (2000) 375 [arXiv:hep-ph/9903282].
  
 \bibitem{Schafer:1996wv}T.~Schafer and E.~V.~Shuryak, Rev.~Mod.~Phys.~\textbf{70}
  (1998) 323 [arXiv:hep-ph/9610451].
  
 \bibitem{Diakonov:1985eg}D.~Diakonov and V.~Y.~Petrov, Nucl. Phys.
  B \textbf{272} (1986) 457.
  
 \bibitem{Diakonov:1995qy} D.~Diakonov, M.~V.~Polyakov and C.~Weiss,
  Nucl.~Phys.~B~ \textbf{461} (1996) 539 [arXiv:hep-ph/9510232].
 
 \bibitem{Goeke:2007bj}
   K.~Goeke, M.~M.~Musakhanov and M.~Siddikov,
     Phys.\ Rev.\  D {\bf 76} (2007) 076007
   [arXiv:0707.1997 [hep-ph]].
   
  
 \bibitem{Dorokhov:2006qm} A.~E.~Dorokhov, W.~Broniowski and E.~Ruiz
  Arriola,   Phys.\ Rev.\ D \textbf{74} (2006) 054023 [arXiv:hep-ph/0607171].
  
 \bibitem{Anikin:2000rq}
   I.~V.~Anikin, A.~E.~Dorokhov and L.~Tomio,
   Phys.\ Part.\ Nucl.\  {\bf 31} (2000) 509
   [Fiz.\ Elem.\ Chast.\ Atom.\ Yadra {\bf 31} (2000) 1023].
 \bibitem{Dorokhov:2003kf}
   A.~E.~Dorokhov and W.~Broniowski,
   Eur.\ Phys.\ J.\  C {\bf 32} (2003) 79
   [arXiv:hep-ph/0305037].
 \bibitem{Kowalski:2003hm}H.~Kowalski and D. Teaney, Phys. Rev. D
  \textbf{68} (2003) 114005 [arXiv:hep-ph/0304189].
  
  
  
  
 \bibitem{Kopeliovich:2007pq}B.~Z.~Kopeliovich, I.~K.~Potashnikova,
  B.~Povh and I.~Schmidt, Phys. Rev. D \textbf{76}, 094020 (2007),
  [arXiv:0708.3636 [hep-ph]].
  
 \bibitem{zeus} S. Chekanov et al., (ZEUS Collaboration), PMC Phys. A1, 6 (2007) [arXiv:0708.1478 ].
  
 \bibitem{Aaron:2007cz} F.~D.~Aaron \textit{et al.} [H1 Collaboration],
   Phys.\ Lett.\ B \textbf{659} (2008) 796 [arXiv:0709.4114 [hep-ex]].
  
  
 \bibitem{Hufner:2000jb} J.~Hufner, Yu.~P.~Ivanov, B.~Z.~Kopeliovich
  and A.~V.~Tarasov,   Phys.\ Rev.\ D \textbf{62} (2000) 094022 [arXiv:hep-ph/0007111].
  
 \bibitem{Kopeliovich:2001xj} B.~Z.~Kopeliovich, J.~Nemchik, A.~Schafer
  and A.~V.~Tarasov,   Phys.\ Rev.\ C \textbf{65} (2002) 035201 [arXiv:hep-ph/0107227].
  
  
 \bibitem{Kopeliovich:1999am} B.~Z.~Kopeliovich, A.~Schafer and
  A.~V.~Tarasov,   Phys.\ Rev.\ D \textbf{62} (2000) 054022 [arXiv:hep-ph/9908245].
   
  
 \bibitem{Iancu:2003ge}E.~Iancu, K.~Itakura and S.~Munier, Phys.
  Lett. B \textbf{590} (2004) 199 [arXiv:hep-ph/0310338].
  
 
 \bibitem{Kowalski:2006hc}H.~Kowalski, L.~Motyka and G.~Watt, Phys.
  Rev. D \textbf{74} (2006) 074016 [arXiv:hep-ph/0606272].
 
\end{thebibliography}
  \end{document}